\documentclass{sig-alternate-2013}
\usepackage{amsmath}
\usepackage{algorithm}
\usepackage[noend]{algpseudocode}
\usepackage{graphicx}
\usepackage{fancyhdr}
\usepackage{textcomp}

\usepackage[
  pass,
]{geometry}

\fancypagestyle{Ackfooter}{%
  \fancyhf{}

  \fancyfoot[L]{\scriptsize{\textbf{ACKNOWLEDGEMENT}: This material is based on work supported in part by the DreamWorks Animation SKG. Any opinions, findings, conclusions or recommendations expressed in this material are the authors\textquotesingle  and do not necessarily reflect those of the sponsors. Additionally, the first author would also like to sincerely thank every member of the NLD group, Dr. Ari Shapiro, Alesia Egan at USC ICT and Prof. Louis-Philippe Morency at LTI, CMU for their insightful suggestions/feedback/co-operation.} }
}

\newfont{\mycrnotice}{ptmr8t at 7pt}
\newfont{\myconfname}{ptmri8t at 7pt}
\let\crnotice\mycrnotice%
\let\confname\myconfname%

\permission{Permission to make digital or hard copies of all or part of this work for personal or classroom use is granted without fee provided that copies are not made or distributed for profit or commercial advantage and that copies bear this notice and the full citation on the first page. Copyrights for components of this work owned by others than ACM must be honored. Abstracting with credit is permitted. To copy otherwise, or republish, to post on servers or to redistribute to lists, requires prior specific permission and/or a fee. Request permissions from permissions@acm.org.}

\begin{document}

\newcommand{\vect}[1]{\mathbf{#1}}
\newcommand{\KGauss}{K_\textit{Gauss}}
\newcommand{\KBern}{K_\textit{Bern}}
\newcommand{\Munct}{\textit{unct}}
\newcommand{\Mdiv}{\textit{div}}
\newcommand{\Mden}{\textit{den}}
\newcommand{\Minfo}{\textit{Info}}

\title{An Active Learning Based Approach For Effective Video Annotation And Retrieval}

\numberofauthors{2}
\author{
\alignauthor{
Moitreya Chatterjee\\
       \affaddr{USC Institute for Creative Technologies}\\
       \affaddr{Playa Vista, CA, USA}\\
       \email{metro.smiles@gmail.com}
}
\alignauthor{
Anton Leuski\\
       \affaddr{USC Institute for Creative Technologies}\\
       \affaddr{Playa Vista, CA, USA}\\
       \email{leuski@ict.usc.edu}
}
}

\maketitle
\begin{abstract}

Conventional multimedia annotation/retrieval systems such as Normalized Continuous Relevance Model (NormCRM)~\cite{lavrenko2004statistical} require a fully labeled training data for a good performance. Active Learning, by determining an order for labeling the training data, allows for a good performance even before the training data is fully annotated. In this work we propose an active learning algorithm, which combines a novel measure of sample uncertainty with a novel clustering-based approach for determining sample density and diversity and integrate it with NormCRM. The clusters are also iteratively refined to ensure both feature and label-level agreement among samples. We show that our approach outperforms multiple baselines both on a recent, open character animation dataset and on the popular TRECVID corpus at both the tasks of annotation and text-based retrieval of videos.

\end{abstract}




\section{Introduction}

The ubiquity of multimedia content in our daily lives requires effective tools for multimedia  annotation and retrieval. Multimedia annotation tools automatically annotate image or video content (samples) with text labels specifying different objects, events, etc. called \textit{concepts}. Most of these systems treat the task of automatic annotation as a classification challenge, whereby a separate classifier is trained for each of these concepts~\cite{lan2014multimedia},~\cite{ghosh2014multimodal},~\cite{benitez2005multimedia},~\cite{park2014computational}. However, fewer approaches explore the correlation between these concepts~\cite{lavrenko2004statistical}.

A typical multimedia retrieval system, on the other hand, ranks the multimedia samples based on their relevance to the user's text query. Generally, the retrieval is done by comparing the query to the sample concept labels. Thus an exhaustive annotation of the sample is often a pre-requisite for such retrieval systems.

Normalized Continuous Relevance Model (NormCRM)~\cite{lavrenko2004statistical} is an example of a technique that allows for a direct retrieval of samples without having to annotate them. However training this model (like many others), requires fully annotated data. The human-effort costs of concept annotation is significant and this raises an interesting research question: is there a way to achieve a decent annotation/retrieval performance without requiring a fully annotated training dataset?

The community has taken to \emph{Active Learning} to address this issue~\cite{huang2008active}. Active Learning, is a machine learning technique that interactively selects unlabeled samples and queries an oracle to provide labels for the samples. Such a system outputs an order of labeling the samples such that a decent annotation/retrieval performance is achieved before all unlabeled data is queried. A typical active learning system consists of a learning engine, which does the annotation/retrieval and a sample selection engine, responsible for determining the labeling order of the unlabeled samples.

In this work, we use NormCRM as the learning engine and propose a novel sample selection algorithm. We call this integrated system CRMActive and apply it for video annotation and video retrieval tasks. The algorithm uses a measure of  \emph{informativeness} for ranking unlabeled samples during active learning. This informativeness combines a new measure of sample uncertainty with a novel cluster-refinement based approach for determining sample density and diversity. Our experiments show that CRMActive outperforms a state-of-the-art approach and a random baseline.

\section{Proposed Approach}

Normalized Continuous Relevance Model (NormCRM) is a generative annotation/retrieval technique~\cite{lavrenko2004statistical}. Let's  consider a video sample $I$ defined by a $M$-dimensional feature vector $\vect{r}$ and $\mathcal{V}$ be the vocabulary of all concept labels (each concept 1 word long). NormCRM defines conditional probability for using a label word $w \in \mathcal{V}$ to annotate the video $I$, as $P(w|\vect{r}) = P(w, \vect{r})/P(\vect{r})$. Lavrenko et al.~\cite{lavrenko2004statistical} suggest that for annotation we pick the top-$k$ words with highest $P(w_i|\vect{r})$, $i=1,2,...,k$. For the task of retrieval using a query word $w$, we pick the top-$t$ videos with highest $P(w|\vect{r}_i)$, $i=1,2,...,t$. In both cases, the joint-distribution of words and features $P(\vect{w}, \vect{r})$ is estimated from the training data by 
\[P(\vect{w}, \vect{r}) = \sum_{J \in T}( P(J) \prod_{w  \in \vect{w}}P(w|J) \prod_{r_i \in \vect{r}, i = 1}^{M}P(r_i|J) ),\]
where $T$ is the set of training video samples and $\vect{w}$ is the set of words in question.

However, NormCRM requires a fully annotated data for training. To circumvent this, we integrate NormCRM into an Active Learning framework by combining it with a sample selection engine, which selects samples for annotation based on their \emph{informativeness}. We calculate the informativeness by combining measures of sample \emph{uncertainty}, \emph{density} and \emph{diversity}. 

\textit{Sample Uncertainty} is a measure of how uncertain the learning engine is about the labels of a sample. Using SVM as a learning engine, entropy and distance of sample from decision boundary have been explored as sample uncertainty measures~\cite{qi2006video, tong2001support}. However, these techniques don't capture a measure of the ambiguity between the relevant labels and the irrelevant ones for NormCRM-based models. Hence, we define a novel measure of uncertainty of an unlabeled sample (defined by a M-dim. feature $\vect{x}$) as:
\begin{equation}
\Munct(\vect{x})=\frac{1}{P(w_{1}|\vect{x}) - P(w_{k+1}|\vect{x})},
\label{eq:uncertainty}
\end{equation}
where $w_{1},...,w_{k}$ (in decreasing order of relevance) are the top-k most relevant labels assigned to $\vect{x}$. The denominator in Eq.~\ref{eq:uncertainty} gives a measure of the gap (distance) between the posterior probabilities of the most relevant label and the first irrelevant one and can thus be used to obtain uncertainty. 

\textit{Sample Density} is a measure of how likely a certain sample is to occur given the underlying distribution that generated the data while a high \textit{Sample Diversity} score ensures that the samples chosen for labeling aren't too similar to each other. To compute sample density and diversity, we start by clustering all samples in the training data $\mathcal{X}=\{\vect{x}_{1}, \vect{x}_{2}, ..., \vect{x}_{N}\}$, consisting of the initial labeled training data $\mathcal{L}$ and the unlabeled training data $\mathcal{U}$ ($\mathcal{X} = \mathcal{L} \cup \mathcal{U}$). We first represent every sample in the visual feature space and perform X-Means clustering. X-Means is a variant of K-Means, which automatically picks the parameter K by comparing the Bayesian Information Criterion (BIC) scores of the clustering system for a range of values of K and picking the one with an optimal score~\cite{pelleg2000x}. We then check if every labeled sample shares a concept with at least one other labeled sample in the same cluster. A sample that shares no labels, is removed from the cluster and we use it to create a new cluster and redistribute unlabeled samples from the original cluster between the old and the new clusters using 2-Means.

In order to measure the extent of agreement amongst the labeled samples in a cluster, both in terms of their visual features and their labels, we use \emph{Empirical Entropy}~\cite{dagli2006leveraging}. For a cluster $C$, it is defined as: 
\begin{equation}
h^{C} =-\frac{1}{n}\sum_{i=1}^{n}\log(\frac{1}{n}\sum_{j=1}^{n}K(\vect{x}_{i}, \vect{x}_{j})),
\label{eq:entropy}
\end{equation}
where there are $n>1$ labeled samples in the cluster and $K(.,.)$ is a kernel function. A kernel is a mapping : $\chi \times \chi \rightarrow \mathbb{R}$, where $\chi$ is the input space. A kernel may be considered as a measure of similarity. For continuous input spaces, such as video features, a Gaussian kernel is often used~\cite{zha2012interactive}:
\[\KGauss(\vect{x}, \vect{x'})=\exp(-||\vect{x}-\vect{x'}||^{2}/2 \sigma ^{2}),\] 
where $\vect{x}, \vect{x'} \in \mathcal{X}$. For discrete input spaces, such as the space of labels, a Bernoulli product kernel may be used~\cite{jebara2004probability}:
\[\KBern(\vect{x}, \vect{x'})=\prod_{d=1}^D[( \gamma _{d}^{x_{d}} \times \gamma _{d}^{x_{d}'}) \times (1 - \gamma _{d})^{(1 - {x_{d}})} \times (1 - \gamma _{d})^{(1 - {x_{d}'})}],\] 
where $\vect{x}, \vect{x'} \in \{0, 1\}^{D}$, $x_{d}$, $x_{d}'$ shows the presence (1) or absence (0) of the  $d^{th}$ concept and $\gamma _{d}$ is the probability of the $d^{th}$ concept occurring. In order to capture the notion of sample similarity both from the visual and label perspectives, we define a new kernel as a combination of the two~\cite{genton2002classes}:
\[K(\vect{x}, \vect{x'})=\KBern(\vect{x}, \vect{x'}) \times \KGauss(\vect{x}, \vect{x'})\]

Once we clustered the sample videos, we compute the sample density of an unlabeled sample $\vect{x}$ in cluster $C$ as
\[\Mden(\vect{x})=\frac{p(\vect{x})}{\max\limits_{\vect{x}_{i} \in \mathcal{X} } p(\vect{x}_{i})},\]
where $p(\vect{x})$ is the kernel density estimate:
\[p(\vect{x})=\frac{1}{|C|}\sum_{\vect{x}_{i} \in C} \KGauss(\vect{x}, \vect{x}_{i})\]
and $|C|$ is the total number of samples in cluster $C$.

\begin{algorithm}
\caption{\normalsize{CRMActive}}\label{CA}
\scriptsize{
\begin{algorithmic}
\State \textbf{Input}: The set $\mathcal{L}=\{\vect{l}_{1}, \vect{l}_{2}, ..., \vect{l}_{P}\}$, their labels $\mathcal{Y}=\{\vect{y}_{1}, \vect{y}_{2}, ..., \vect{y}_{P}\}$ where $\vect{y}_{i} \in \{0, 1\}^{D}$, the set $\mathcal{U}=\{\vect{u}_{1}, \vect{u}_{2}, ..., \vect{u}_{Q}\}$ and $K$ :- nos. of samples to pick in a batch.
\State \textbf{Output}: The set $\mathcal{L}$, containing the order in which the unlabeled samples are labeled.
\State \textbf{Algorithm}:
\State Perform X-Means, using the visual features, on the set of $\mathcal{L} \cup \mathcal{U}$ samples. Say, T be the optimal number of clusters and let \textit{rep}($C_{i}$) denote the representative sample of cluster $C_{i}$.
\State Check if $\forall \vect{l}_{j}, \vect{l}_{j} \in \mathcal{L}, \vect{l}_{j} \in C_{k}, \vect{l}_{j}$ shares $\geq$ 1 concept with at least 1 labeled sample in $C_{k}$, otherwise call Redistribute($C_{k}, \vect{l}_{j}$).
\State $h_{worst} := $ NIL // Initialize $h_{worst}$
\While {$\mathcal{U} \neq \phi$}
\State Train NormCRM using $\mathcal{L}$, evaluate model on test set.
\State Update $h_{worst}$ to max. entropy value among all clusters with at least 2 labeled samples
\State Compute \textit{Info}$(\vect{x}_{i}), \forall \vect{x}_{i} \in \mathcal{U}$
\State Pick top-$K$ samples, $\vect{Lab} = \{\vect{a}_{1}, \vect{a}_{2}, ..., \vect{a}_{K}\}$ for labeling. 
\State $\mathcal{L} := \mathcal{L} \cup \vect{Lab} , \mathcal{U} := \mathcal{U} - \vect{Lab}$ // Update the lists
\State // Now refine the clusters based on newly labeled samples
\For {$j = 1, 2, ...,$ K}
\If {$h_{worst} =$ NIL} // If $h_{worst}$ is not set
\State Check if sample $\vect{a}_{j}, \vect{a}_{j} \in C_{k}$ shares $\geq$ 1 concept with at least 1 labeled sample in $C_{k}$, otherwise call Redistribute($C_{k}, \vect{a}_{j}$).
\Else // Determine which sample in $C_{k}$ to knock out
\State Compute $h^{C_{k}}$, where $\vect{a}_{j} \in C_{k}$ //$C_{k}$ > 1 labeled sample
\If {$h^{C_{k}} > h_{worst}$} // Exceeds threshold
\For {$r = 1, 2, ...,$ \# labeled samples in $C_{k}$}
\State $C_{k}' :=  C_{k} - r^{th}$ labeled sample in $C_{k}$
\If {$h^{C_{k}'} \leq h_{worst}$} // Meets threshold
\State $\vect{W} := r^{th}$ labeled sample
\State Redistribute($C_{k}, \vect{W}$) // Split cluster
\State break
\EndIf
\EndFor
\EndIf
\EndIf
\EndFor
\EndWhile
\end{algorithmic}
}
\end{algorithm}

Our definition of the sample density, though similar to Zha et al.~\cite{zha2012interactive}, differs by using clusters, which are refined (see later in this section), to determine the neighboring samples of $\vect{x}$ rather than a static set of its k-nearest neighbors. 

\begin{algorithm}
\scriptsize{
\begin{algorithmic}
\Procedure{Redistribute}{Samples in $C_{k}$, $\vect{a}$}
\State \textbf{Input}: Set of all samples in cluster $C_{k}$ \& the seed sample $\vect{a}$
\State \textbf{Output}: Updated set of clusters
\State \textbf{Algorithm}:
\State Create a new cluster, $C_{k}'$,  with $\vect{a}$ as the centroid.
\State Perform \textbf{2-Means} on the unlabeled samples of cluster $C_{k}$ with \textit{rep}($C_{k}$) and $\vect{a}$ as the two initial cluster centroids.
\State Update \textit{rep}($C_{k}'$) as the representative sample of cluster $C_{k}'$.
\State Determine the centroid of the labeled and the remaining unlabeled samples in $C_{k}$ and similarly update \textit{rep}($C_{k}$). 
\EndProcedure
\end{algorithmic}
}
\end{algorithm}

To compute the sample diversity, we use the angular distance between features similar to Brinker's technique~\cite{brinker2003incorporating}. However we choose only the representative samples of every cluster (i.e. the sample closest to the cluster centroid), $rep(C)$, rather than all the samples in $\mathcal{X}$, to gain speed. Diversity of the unlabeled samples is thus, defined as: 
\[\Mdiv(\vect{x}) = 1 - \max_{\vect{x}_{i} \in \mathcal{S}} \frac{\KGauss(\vect{x}, \vect{x}_{i})}{\sqrt{\KGauss(\vect{x}, \vect{x})\times\KGauss(\vect{x}_{i}, \vect{x}_{i})}},\]
where $\mathcal{S}$ is the set of all $T$ cluster representatives $\mathcal{S}=\{rep(C_{1}), rep(C_{2}), ..., rep(C_{T})\}$ .

Now, we combine these measures to determine the \emph{informativeness} of an unlabeled sample $\vect{x}$ as 
\[\Minfo(\vect{x}) = \lambda _{1} \times \Munct(\vect{x}) + \lambda _{2}  \times \Mden(\vect{x}) + \lambda _{3} \times  \Mdiv(\vect{x}).\] 

We rank the unlabeled samples in the order of decreasing $\Minfo(\vect{x})$ score, to select a batch of top-$K$ samples for labeling. While Zha et al. use a combination of sample local structure, density, diversity, and relevance to score the samples~\cite{zha2012interactive}, our approach differs, most notably, in the use of clustering and a novel uncertainty measure.

Equation~\ref{eq:entropy} reveals that a cluster with low inter-sample disagreement has a low entropy. As more samples in a cluster $C$ are labeled, the disagreement among its labeled samples increases. This changes the empirical entropy $h^{C}$ in a monotonically non-decreasing fashion. Therefore we refine the clusters by doing the following: After each batch of labeling, the algorithm determines the cluster with the worst entropy and uses its $h^{C}$ as a threshold to decide whether to keep or split a cluster during the next batch and this is repeated for successive iterations. If a newly labeled sample increases the cluster entropy beyond the threshold for that batch, then a grid search is used to determine the first labeled sample without which the cluster meets the entropy threshold. We create a new cluster with this sample and rearrange the unlabeled samples via 2-Means, like before (see Algo.~\ref{CA}).

\section{Experiments}

We conduct two sets of experiments. In each set, the experimental dataset is divided into training and test subsets. For the first set of experiments, the task of an algorithm is to annotate a test video with a subset of concepts from the vocabulary. The algorithm starts with the training data set divided into labeled ($\mathcal{L}$) and unlabeled ($\mathcal{U}$)  parts. Initially only a small subset of the training set is considered to be labeled. The algorithm uses this information to annotate the test set with concept labels. For the next step, the algorithm selects a batch of $K$ unlabeled training samples, we reveal the labels for the selected samples, and the algorithm repeats the annotation task. For every iteration, we compute precision scores of the algorithm on the test-set for each concept and report their average. We call this score: AP.

In the second set of experiments, an algorithm ranks the test samples by their similarity to a single word query without annotating the test samples. Again, the algorithm starts with the training dataset divided into labeled and unlabeled parts. For each concept label in the vocabulary, the algorithm ranks the test samples by their similarity to the concept. It then selects a batch of $K$ unlabeled training samples, we reveal the labels for the selected samples, and the algorithm repeats the ranking task. For each round, we report the AP scores for the top 5 images/videos.

\subsection{Datasets}

\textbf{TRECVID 2007}: The TRECVID 2007 video corpus has 110 short video clips~\cite{TRECVID07}. Each frame in every video is annotated with at most 16 concept labels selected from a set of 36 concepts such as ``crowd'', ``building'', ``airplane'', etc. This corpus has been used extensively in video annotation experiments~\cite{zha2012interactive}. In recent multimedia recognition/annotation tasks histograms have been found to be effective as a feature summarization technique for text content(~\cite{joachims2002learning},~\cite{chatterjee2014verbal},~\cite{shim2015acoustic}), acoustic content(~\cite{kuhn1990cache}) and images/video(~\cite{chatterjee2014context},~\cite{zha2012interactive},~\cite{dalal2005histograms}). Therefore, for every frame we compute a 225-dimensional feature vector (color moment, edge orientation histogram, wavelet PWTTWT texture) as described in the work of Zha et al.~\cite{zha2012interactive}. We test our model on the frames from 13 randomly selected videos and we use the rest of the data (frames from 97 videos) for training. We selected 4000 frames from the training data as the initial set of labeled samples $\mathcal{L}$, containing at least 1 positive example of every concept. We set, batch size, $K$ to 2400.

\textbf{USC SmartBody}: SmartBody is an open virtual character animation platform. It ships with a library of 274 animations such as walking, hand beat gesture, pointing, eye-brow raising, lip corner stretching, etc.~\cite{thiebaux2008smartbody}. The animations are defined on a 3D skeleton consisting of 119 individual joints and the 3D coordinates of these joints are available from the SmartBody API. Each animation is annotated using at most 6 concept labels from a set of 30 labels such as ``Legs'', ``Arms'', ``Face'', ``Left'', ``Right'', etc. The X-axis of Figure~\ref{ConceptAP} gives an exhaustive list of all the concepts. The animations are annotated at the video clip level (i.e. the individual frames are not annotated). 9 out of 119 joints have been handpicked for feature computation (neck, left(L)/right(R) shoulders, L/R elbows, L/R hip joints, and L/R knees). For each frame in an animation, the skeleton angles at these joints are computed~\cite{sedmidubsky2013retrieving} and the differences between the minimum and the maximum values for the angles during the whole animation sequence have been encoded as a 9-dimensional feature vector. This dataset called the USC SmartBody Annotation-Retrieval Dataset (SARD) has recently been made available for research by the community~\cite{chatterjee2015crmactive}. We randomly selected 24 animations for testing and we use the rest of the data (250 animations) for training. We selected 40 animations from the training data as the initial set of labeled samples $\mathcal{L}$, containing at least one positive example of each concept. We now set, batch size, $K$ to 23.

\subsection{Baseline Systems}

For annotation task, we compare CRMActive with two methods. The first one is an active learning system that uses NormCRM as the learning engine while the samples are selected randomly. The results are averaged over 3 runs with different random seeds. The second baseline is the method proposed by Zha et al. (state-of-the-art)~\cite{zha2012interactive}. We determine the two NormCRM smoothing parameters $\lambda$ and $\beta$~\cite{lavrenko2004statistical}, and the validated parameters of the second baseline using 10-fold cross-validation on the first annotation batch. These values are then fixed for successive rounds. The values of the fixed parameters for the second baseline are reused from the paper~\cite{zha2012interactive}. For CRMActive, probability $\gamma _{d}$, is re-estimated from the labeled training data on each annotation batch and the weighting parameters $\lambda _{i}=\frac{1}{3}, i = 1..3$. Finally, both NormCRM and CRMActive work by ranking annotation concepts, so we assign the top-16 concepts for TRECVID 2007 and the top-6 for SmartBody as relevant. For direct retrieval, CRMActive is compared only with the first baseline discussed above, since no prior work is known.

\begin{table}[htbp]
   \centering
   \begin{tabular}{@{} c|c @{}} 
 \hline 
 \includegraphics[scale=.40]{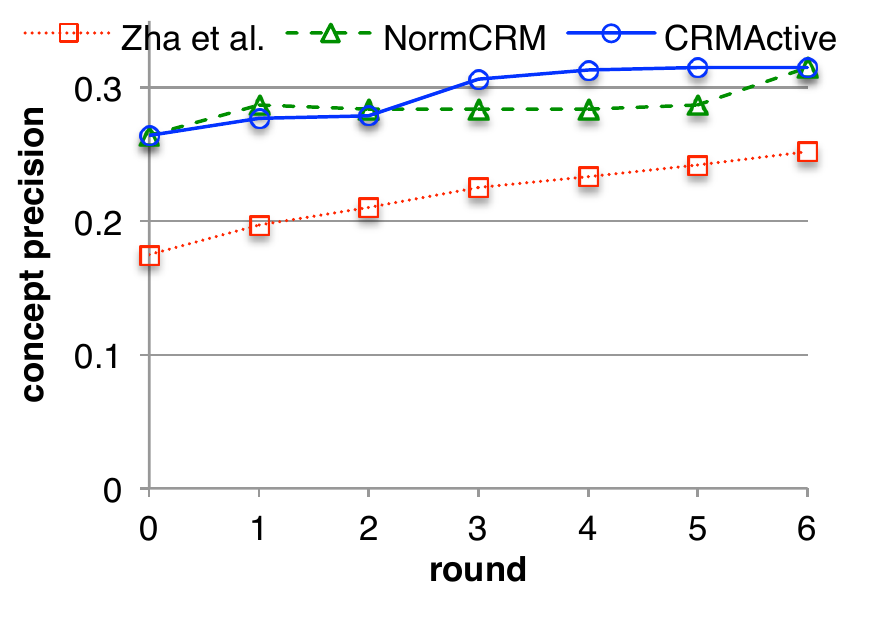}  & \includegraphics[scale=.40]{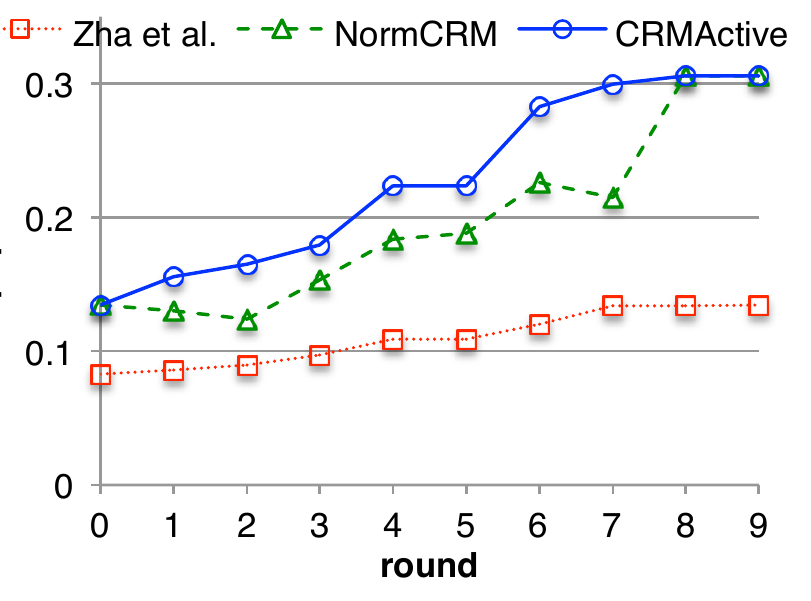} \\
 (a) & (b) \\
 \hline 
 \includegraphics[scale=.40]{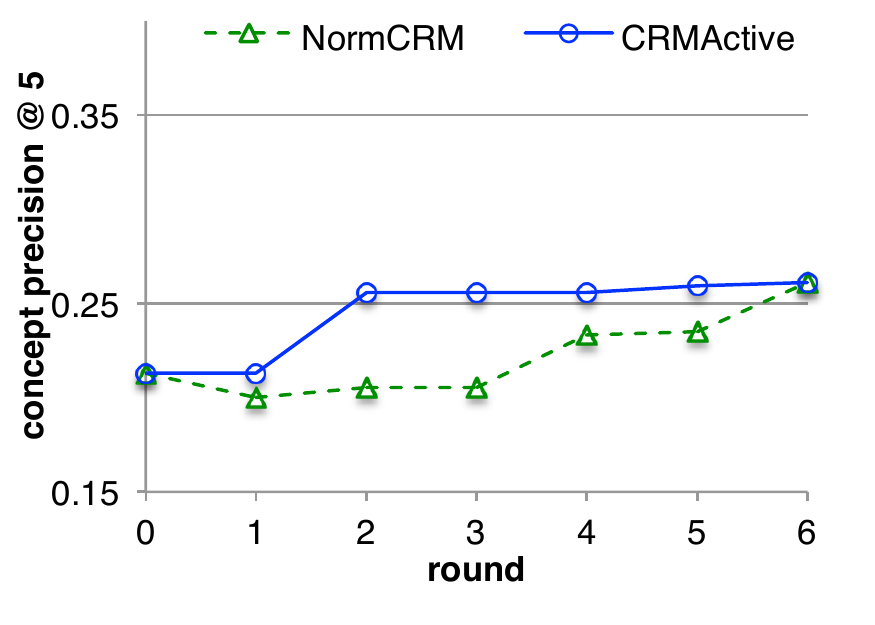} & \includegraphics[scale=.40]{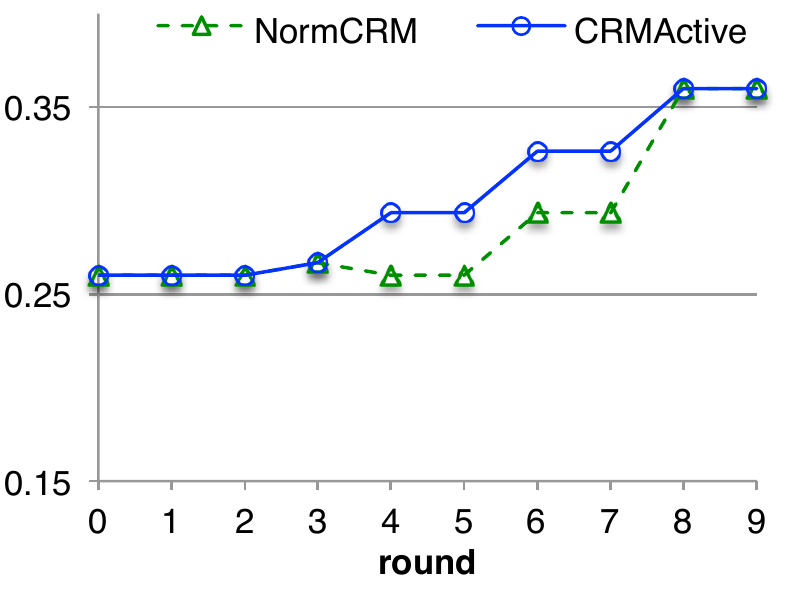} \\
 (c) & (d) \\
 \hline 
   \end{tabular}
   \caption{AP scores (on Y-axis) for annotation on TRECVID (a), SmartBody (b) and AP scores (on Y-axis) for retrieval of top-5 videos on TRECVID (c), SmartBody (d).}
   \label{APs}
\end{table}

\subsection{Results and Discussion}

\begin{figure}
\centering
\includegraphics[width=230pt, height=95pt]{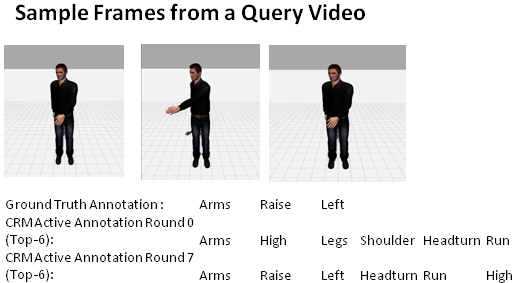} 
\caption{\normalsize{A sample annotation result on SmartBody dataset, showing the top-6 annotated labels by CRMActive after Round 0 and Round 7.}}
\label{SampRes}
\end{figure}

The results in Table~\ref{APs} shows that both the NormCRM-based models, i.e. the first baseline (NormCRM) and CRMActive, generally perform better than the Zha et al. approach for annotation. We believe that this is due to the fact that NormCRM captures the inter-label correlation while Zha et al. trains individual classifiers for every concept. Also the NormCRM-based systems jointly model the labels and features, which allows them to capture the patterns from both these perspectives, this is again not the case for Zha et al. Furthermore CRMActive by selecting the more informative samples first, trains a more robust model early on, which results in its monotonic non-decreasing AP score for annotation/retrieval. This is in contrast with the occasional dips in the AP scores of the random baseline, which might potentially select some of the relatively ``bad" (noisy) training samples early on. Figure~\ref{SampRes} shows a sample annotation result on the SmartBody dataset using CRMActive. We see that the model gets all top 3 labels correct at Round 7, even before the training data is fully annotated.

\begin{figure}[t]
\centering
\includegraphics[scale=0.55]{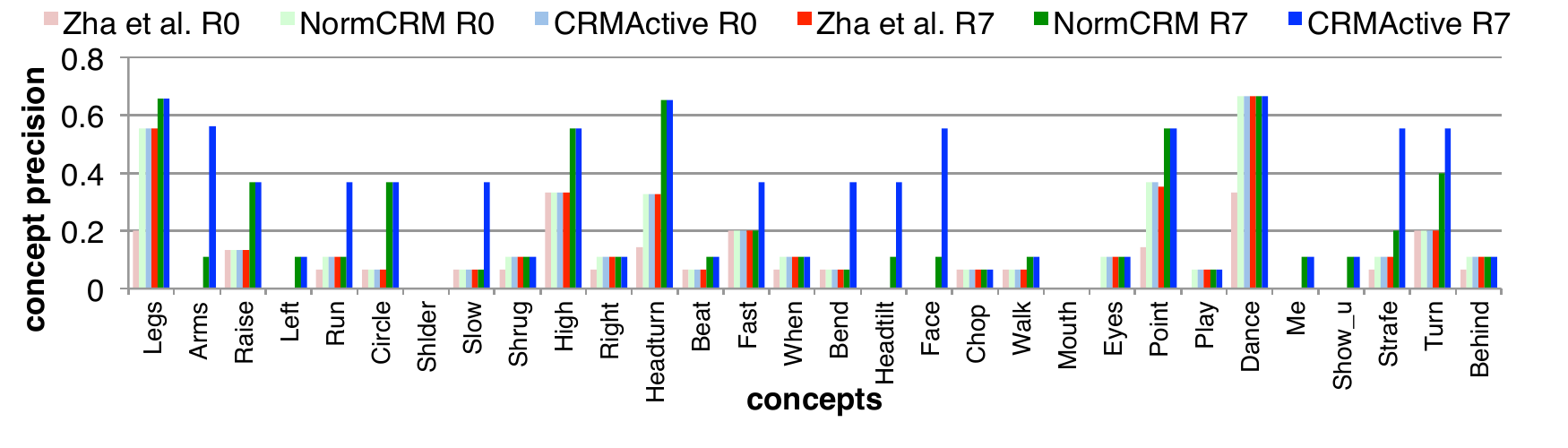} 
\caption{Precision scores for annotation of individual concepts of SmartBody for Round 0 (R0) and Round 7 (R7) of active learning.}
\label{ConceptAP}
\end{figure}

Figure~\ref{ConceptAP} shows the annotation performance of all the models for the individual concepts of the SmartBody dataset over two rounds (initial and towards the end). The concept scores for the NormCRM random baseline are obtained by averaging over the results of the 3 runs. We notice a performance gain for all the models across most concepts over the two rounds, indicating that more training data helps. We also notice that CRMActive is always at least as good, on all concepts. For concepts with a high number of positive examples, such as Legs, all models do well.  Further, we believe that the nature of the features used can explain a good performance by all models for complex concepts such as Dance as compared to some others ones like Mouth.


\section{Conclusions}
In this work, we proposed a sample selection algorithm based on active learning by combining a novel measure of sample uncertainty and a novel cluster-refinement approach for determining sample density and diversity. This approach is shown to outperform multiple baselines at both annotation and retrieval tasks. Our experiments also reveal the pros of using a generative approach of jointly modeling both the features and labels. CRMActive is thus shown to be a promising active learning approach to explore.

\textbf{ACKNOWLEDGEMENTS:}

This material is based on work supported in part by the DreamWorks Animation SKG. Any opinions, findings, conclusions or recommendations expressed in this material are the authors\textquotesingle  and do not necessarily reflect those of the sponsors. Additionally, the first author would also like to sincerely thank every member of the NLD group, Dr. Ari Shapiro, Alesia Egan at USC ICT and Prof. Louis-Philippe Morency at LTI, CMU for their insightful suggestions/feedback/co-operation.


%
\small{
\bibliographystyle{abbrv}
\bibliography{Paper_Version_Cam2} 
} 

%
%

\end{document}